\documentclass[final,floatfix,twocolumn,pre,aps,floats,amsfonts,amssymb,showpacs]{revtex4}

\usepackage{graphicx}
\usepackage{bm}

\def\EQ{\begin{equation}}
\def\EN{\end{equation}}
\def\EQA{\begin{eqnarray}}
\def\ENA{\end{eqnarray}}

\begin{document}
\title{
Momentum transport and torque scaling in Taylor-Couette flow from an 
analogy with turbulent convection}
\author{ B. Dubrulle$^{1}$ and F. Hersant$^{2,3}$}

\address{
$^1$ CNRS, Groupe Instabilit\'e et Turbulence, CEA/DRECAM/SPEC, 
F-91191 Gif sur Yvette Cedex, France\\
$^2$ CNRS, Service d'Astrophysique, CE Saclay, F-91191 Gif sur Yvette 
Cedex, France\\
$^3$ CNRS, UMR 8632, D\'epartement de Recherches Spatiales, Observatoire de Meudon, 
F-92195 Meudon Cedex, France
}

\begin{abstract}
  We generalize an analogy between rotating and stratified shear flows. This analogy is summarized in Table 1. We use this analogy in the unstable case (centrifugally unstable flow v.s. convection) to 
compute the torque in Taylor-Couette configuration, as a function of 
the Reynolds number. At low 
Reynolds numbers, when most of the dissipation comes from the mean 
flow, we predict that the non-dimensional torque $G=T/\nu^2L$, where $L$ is the cylinder length, scales with 
Reynolds number $R$ and gap width $\eta$,
$G=1.46 \eta^{3/2} (1-\eta)^{-7/4}R^{3/2}$. At larger Reynolds 
number, velocity fluctuations become non-negligible in the 
dissipation. In these regimes, there is no exact power law dependence 
the torque versus Reynolds. Instead, we obtain logarithmic 
corrections to the classical ultra-hard (exponent $2$) regimes:
$$
G=0.50\frac{\eta^{2}}{(1-\eta)^{3/2}}\frac{R^{2}}{\ln[\eta^2(1-\eta)R^ 
2/10^4]^{3/2}}.$$
These predictions are found to be in excellent agreement with 
available experimental data. Predictions for scaling of velocity 
fluctuations are also provided.
\end{abstract}


\date{Eur. Phys. J. B vol 26 p 379-386 (2002)}

\maketitle

\narrowtext
\section{Motivation and objectives}

At sufficiently large Reynolds number, the fluid between co-rotating 
coaxial cylinders becomes turbulent, and a significant momentum 
transport occurs between the two cylinders. In the case with rotating 
inner cylinder and resting outer one (the so-called Taylor-Couette 
flow), detailed measurements show that the torque applied at 
cylinders by the turbulent flow is a function of the Reynolds number 
$R$. There is no clear consensus about this dependence yet: marginal 
stability computation of King et al \cite{KLSM84} or  Barcilon and 
Brindley \cite{BarcBrin84} predict that the non-dimensional torque 
$G=T/\nu^2L$, where $L$ is the cylinder height should vary like $G\sim R^{5/3}$. Old experimental data indicated 
the existence of two scaling regimes, one for $R>10^4$ where the 
exponent is $1.5$, and one for larger Reynolds number, where the 
exponent switches towards $1.7-1.8$ \cite{Wendt33,Taylor36,TGHW90}.  
Recent high precision experimental data yielded no region of constant 
exponent, and revealed a transition with a marked change of 
approximate slope of $G$ as a function of $R$ 
\cite{LFS92,LewiSwin99}. This observation led Eckhardt et al 
\cite{EGL00} to propose a new theory, in which the dependence $G$ 
versus $R$ is through a superposition of scaling laws (describing 
contribution from a boundary layer and the bulk flow). They claim 
that this superposition fits the data better than the Prandtl-Karman 
skin friction law proposed by \cite{Pant92,LFS92,LewiSwin99}. Note that all the scalings are within the theoretical bound derived by Doering and Constantin \cite{DoerConst92}, which implies that the non-dimensional torque cannot increase faster than $R^2$.\

The observational features are reminiscent of heat transport in 
turbulent thermal convection, where approximate scaling laws and 
transition between different regimes have also been observed (for a 
review see \cite{GrosLohs00}). In fact, this similarity is pointed 
out in \cite{LFS92}, and in \cite{EGL00}, and similar techniques are 
used in \cite{MalkVero58,KLSM84} and in \cite{GrosLohs00,EGL00} to 
derive theoretically the scaling regimes in the Rayleigh-B\'enard 
system, and in the Taylor-Couette system. However, the similarity is 
more than superficial: as well known since several decades 
\cite{Brad69}, there is an exact analogy between equations of motions 
of rotating and stratified shear flows (stable or not). There must therefore exist an 
exact analogy between the momentum and heat transport in these two 
systems, although it has so far never been explored. Our goal here is 
to derive this analogy, and examine its consequences in the unstable regime, where the angular momentum or temperature stratification leads to a linear instability. We thus in this paper mainly focus on the analogy between centrifugally unstable Taylor-Couette flow, and convection. 

\section{The analogy}
\subsection{Reminder}
The root of the analogy can be found in the Lamb formulation of the 
Navier-Stokes equations:
\EQ
\partial_t {\bf u}-{\bf u}\times {\bf \omega}=-\nabla\left (
p+\frac{u^2}{2}\right)
+\nu\Delta {\bf u},
\label{rootine}
\EN

where ${\bf \omega}$ is the vorticity, $p$ is the pressure and $\nu$ the 
molecular viscosity. The constant density has been set equal to one 
for simplicity.
In a rotating shear flow ${\bf u}=V(r) {\bf e_\theta}$, the vorticity is 
only in the axial direction and the Lamb vector ${\bf u}\times {\bf 
\omega}$ acts only in the radial direction. Its contribution can be 
split in two parts:
\EQ
{\bf e_r} \cdot ({\bf u}\times {\bf \omega})=VS+2\frac{\Omega}{r} {\cal L},
\label{twoterms} 
\EN
where ${\cal L}=rV$ is the angular momentum, $\Omega=V/r$ is the angular velocity and
$S=r\partial_r\Omega$ is the shear. The first contribution is the 
exact analog of the Lamb vector of a pure shear flow, in a plane 
parallel geometry. The second contribution reflects the stabilizing 
influence of the Coriolis force. Its analog would be produced by 
temperature stratification in the spanwise direction of a planar 
shear flow. The equation of angular momentum conservation then 
suggests to split further this analogy by requiring that 
$2\Omega/r\sim \beta g$, where $g$ is the gravity, and $\beta$ a 
coefficient of thermal expansion, and ${\cal L}\sim \Theta$, the potential 
temperature. This remark is at the heart of the analogy between the 
stability properties of rotating and stratified shear flows, and has 
been used in the past (see e.g. \cite{Brad69}...). Our point here is to 
show that it can be extended into the turbulent regime, via a new 
Langevin model of small-scale turbulence. This new model is based on Rapid Distorsion Theory, i.e. on linearized equations for the small-scale motions. This linear structure explains the possibility of extension of the analogy towards the turbulent regime.

\subsection{The turbulent model}

The turbulent model has been described and tested in \cite{LDN01} for 
general 3D flows, in \cite{NKD00,DLNK01} for shear flows and in 
\cite{DLSW02,DLS02,Dubr00,Dubr01} for stratified shear flows.
In this model, the dynamics of the turbulent flow is obtained from 
solutions of two coupled sets of equations. The first one described 
the dynamics of the mean  velocity ${\bf U}$:
\EQ
\partial_t U_i+\partial_j U_i U_j+\partial_j<u_i' u_j'>
=-\partial_i P
+\nu\partial_j\partial_j U_i,
\label{LSEq}
\EN
Here, the primes denote fluctuating quantities and $<>$ 
the averaging.
To close the system, we need $<u_i'u_j'>$. They are obtained as 
solution of a linear, stochastic equation valid for localized 
wave-packets of velocity and temperature:
\EQA
D_t \hat u_i
&=&-ik_i \hat p- \hat u_j \partial_j U_i
-\nu_t k^2 \hat u_i+\hat f_i\nonumber\\
k_i \hat u_i
&=&0,
\label{ssEqmod}
\ENA
where
\EQ
{\hat u}({\bf x},{\bf k},t)=\int g(\vert
{\bf x-x'}\vert)e^{i{\bf k\cdot (x-x')}}
{\bf u}({\bf x'},t)d{\bf x'},
\label{gabordef}
\EN
$g$ being a function which decreases rapidly at infinity.
We have dropped primes on fluctuating quantities for convenient notations
and introduced the total derivative $D_t=
\partial_t +U_j\partial_j
-\partial_j (U_i k_i) \partial_{k_j}$ . Once the solutions of (\ref{ssEqmod}) have been computed, the Reynolds stress is found by an inverse Gabor transform as:
\EQ
<u'_i u'_j>=\int d{\bf k} \left( u'_i({\bf k},{\bf x}, t) u'_j(-{\bf k},{\bf x},t)+ ( u'_i(-{\bf k},{\bf x}, t) u'_j({\bf k},{\bf x},t)\right).
\label{inverse}
\EN 
Note that the linear part of (\ref{ssEqmod}) is exact and describes 
non-local interactions between the mean and the fluctuating part. The 
major approximation of the model is to lump the non-linear terms 
describing local interactions between fluctuations into a turbulent 
viscosity $\nu_t$. The force $f$ appearing in (\ref{ssEqmod}) is a 
small scale random forces which is introduced to model the seeding of 
small scales by energy cascades (for example via turbulent 
structures, detaching from the wall).\

\subsection{The Taylor-Couette case}
In the Taylor-Couette (rotating shear flow) case, the equations for the azimuthal component 
of the velocity $V(r)$ simplify into:
\EQ
\partial_t V+\frac{1}{r^2}\partial_r r^2<uv>
=\nu\left(\nabla^2 V-\frac{V}{r^2}\right).
\label{afinir}
\EN
The equation for the fluctuations $(u,v,w)$ become:
\EQA
D_t \hat u&=& 2\frac{k_r k_\theta}{k^2}\left(\Omega+S\right) \hat 
u+2\Omega\hat v\left(1-\frac{k_r^2}{k^2}\right) -\nu_t k^2 \hat 
u+\hat f_r,\nonumber\\
D_t \hat v&=& 2\frac{k_\theta^2}{k^2}\hat u (\Omega+S)-2\frac{k_r 
k_\theta}{k^2}\hat v\Omega-\left(2\Omega+S\right) \hat u-\nu_t k^2 
\hat v+\hat f_\theta,\nonumber\\
D_t \hat w&=& 2\frac{k_\theta k_z}{k^2}\hat u (\Omega+S)-2\frac{k_r 
k_z}{k^2}\hat v\Omega-\nu_t k^2 \hat w+\hat f_z.
\label{fluctu}
\ENA
Here, we have used the incompressibility to eliminate the pressure. 
These equations have to be supplemented by the equations describing 
the ray trajectories:
\EQA
\dot{r}&=&0,\quad \dot{\theta} =\Omega,\quad \dot{z}=0,\nonumber\\
\dot{k_r}&=&-k_\theta S,\quad \dot{k_\theta}=0,\quad \dot{k_z}=0.
\label{raytraj}
\ENA

We now introduce a pseudo-temperature
\EQ
\hat \theta=r(\hat v-\frac{k_\theta}{k_z}\hat w)=i r\frac{\hat \omega_r}{k_z},
\label{pesuedotemp}
\EN
where $\hat\omega_r$ is the radial vorticity. With this temperature 
and using the incompressibility condition ${\bf k}\cdot {\bf u}=0$,
we can rewrite (\ref{fluctu}) as:
\EQA
D_t \hat u&=& 2 S \frac{k_r k_\theta}{k^2}\hat 
u+2\frac{\Omega}{r}\frac{k_z^2}{k^2}\hat\theta -\nu_t k^2 \hat 
u+\hat f_r,\nonumber\\
D_t \hat \theta&=& (2\Omega+S)r \hat u -\nu_t k^2 \hat 
\theta+\hat f_\theta,\\
\label{analogb}
\ENA
The set of equation (\ref{analogb}) is the exact analog of the 
equations describing the behavior of vertical and temperature fluctuations in a stratified 
shear flow (see \cite{DLSW02,DLS02,Dubr00,Dubr01} for their 
expression), provided the correspondence summarized in Table 1. 
holds. Note that the analog of 
the temperature is not the angular momentum, but related to the 
z-integral of the radial vorticity (in Gabor variable, integration on 
z is done via division by $k_z$). At large scale, since the velocity 
profile is axi-symmetric, this integral of the radial vorticity 
reduces to the angular momentum, as previously suggested 
\cite{Brad69}.\

\subsection{Stability and Importance of axi-symmetric modes}

The correspondence described in Table 1. generalizes the well-known 
analogy established previously \cite{Brad69} for the stability 
analysis under axi-symmetric perturbations (case where $k_\theta=0$). 
In particular, from (\ref{analogb}), one can write a compact differential equation for $\omega=u k^2$ :
\EQ
D_t^2 \omega +\kappa^2 k_z^2\frac{\omega}{k^2}-\nu k^2 \omega +f_\omega=0,
\label{bradshaw}
\EN
where 
$\kappa^2=2\Omega(2\Omega+S)$ is the epicyclic frequency. Using the ray equation 
(\ref{raytraj}), one can then define a non-dimensional number $B=\kappa^2 k_z^2/(k_\theta^2+k_z^2)$ (the Bradshaw number) which governs the stability of the wave packet along the trajectory. This number is the analog of Richardson number in stratified shear flow. For example, it can be shown that in the absence of diffusion, the amplitude of the wavepacket has a monotonic (growing for one mode, decaying for another one) behavior at late time for $B<1/4$, while it becomes oscillatory for $B>1/4$. Clearly, the oscillatory behavior creates dephasing effects for the Reynolds stresses, which may lead to its pure cancellation, thereby removing the influence of the small scales onto the large scale. We therefore identify the regime with $B>1/4$ as a regime with purely laminar motions, where turbulence effects are strongly suppressed.  This property tends to favor bi-dimensional modes (those for which $k_z=0$) since in this case the epicyclic frequency can take any value for non-oscillatory behavior. The inclusion of diffusion changes the mode selection. One 
can indeed check that the viscous decay is proportional to $\exp(-R^{-1} 
tS)$ rather than $\exp(-R^{-1}(tS)^3)$ for non-axi-symmetric 
perturbations. This shows that axi-symmetric perturbation (with $k_\theta=0$) are favored with respect to non-axi-symmetric perturbation. In this case, the Bradshaw number becomes independent of the wavenumber of the wave packet, and one can identify a new boundary of stability according to its sign: when it is positive, axi-symmetric perturbations can be exponentially amplified and superseed viscous decay. We call this regime "unstable". It is the analog of the convective regime in the stratified case. In the sequel (Section 3), we shall concentrate on this regime, leaving the other regime for further study. 

\subsection{Completion of the analogy in the unstable case}

For axi-symmetric modes, $\omega_r/k_z=v$ and the equation for the mean angular momentum ${\cal L}$ can then be written in equivalent form:
\EQ
\partial_t {\cal L}+\frac{1}{r}\partial_r (r<u\theta>)=\nu\left(\nabla^2 {\cal L}/r-\frac{{\cal L}}{r^2}\right).
\label{analogfinal}
\EN
Comparing this equation with the equation giving the mean temperature profile, we 
finally remark that the only difference lies in the viscous terms, 
because in cylindrical coordinates, the Laplacian includes terms 
describing curvature effects. In the most general case, this forbids 
the analogy to be drawn at the level on mean profile (ie after 
integration over $r$ of eq. (\ref{afinir}): for example, it is well 
known that in stratified shear flow, the laminar temperature profile 
is linear, while its analog, the laminar angular momentum profiles 
varies like: $L\sim Ar^2+B$. In many Taylor-Couette experiments, 
however, the gap between the two cylinders is small, and curvature 
effects can be neglected. One can for example check that the angular 
momentum in the experiments by \cite{LFS92} is linear in the laminar 
regime, while it flattens at the center of the gap in the turbulent 
regime, exactly like its temperature analog. In the sequel, we shall 
assume a small gap geometry, and neglect curvature effects.

\section{Application of the analogy in the unstable case}

The present analogy is the turbulent generalization of a previously 
known analogy for axi-symmetric modes. In the sequel, we shall use previous considerations about stability of axi-symmetric modes to assume that the turbulent properties are dominated by the contribution of the axi-symmetric modes, i.e. restrict ourselves to these modes. The relevence of this approximation will be tested by comparisons of its predictions regarding some characteristic quantities measured, in the 
turbulent regime with experimental data. 

\subsection{Stability}

In the unstable regime, 
a classical parameter describing the intensity of the convection is 
the Rayleigh number:
\EQ
Ra=\frac{\beta gD^3\Delta \Theta}{\kappa\nu},
\label{Rayleigh}
\EN
where $D$ is the size of the cell in the stratified direction, 
$\Delta \Theta$ is the temperature gradient applied.
In most convection experiments, this number is unambiguously defined 
because of the constancy of the gravity at the scale of the 
experiment. In the Taylor-Couette case, the gravity depends on the 
perturbation, and one may wonder how to define this Rayleigh number 
in a general way. In a recent analysis of stability of 
Taylor-Couette experiment, Esser and Grossman \cite{EsseGros96} 
suggested to evaluate this factor at the gap center, 
$r_c=(r_1+r_2)/2$, leading to
\EQA
Ra_*&\equiv&-2\frac{\Omega}{r}\partial_r 
{\cal L}\frac{d^4}{\nu^2}\vert_{r=r_c},\nonumber\\
&=&4\frac{\eta^2}{(1-\eta^2)^2}\left(\frac{d^2}{r_c^2}-\frac{d^2}{r_2^2 
}\right)R^2,\nonumber\\
&=&4\frac{\eta^2(1-\eta)(3+\eta)}{(1+\eta)^4}R^2.
\label{rayleighequiv}
\ENA
In the sequel, we shall use a star label to refer to analog 
quantities. In (\ref{rayleighequiv}), we have used $\kappa_*=\nu$ and 
introduced the Reynolds number $R=r_1 d\Omega_1/\nu$, where $r_1$ is 
the internal radius, $\Omega_1$ the rotation rate at the inner 
radius, $d$ the gap width and $\eta=r_1/r_2$.
Note that the analog Rayleigh number $Ra_*$ varies with the radial 
aspect ratio $\eta$. In the small gap approximation $\eta\to 1$, 
experiments show that the critical Rayleigh number tends to a 
constant $Ra_*\approx 1706$. This value is very close to the value 
$Ra=1707.762$ obtained in Rayleigh-B\'enard convection for rigid 
boundary conditions \cite{Chan70}. In sheared convection, the 
critical Rayleigh number is modified with respect to this theoretical 
value, and display corrections quadratic in the Reynolds number based 
on the shear. In the present case, these correction are proportional 
to $(1-\eta)^4\to 0$, and the critical Rayleigh number stays close to 
the un-sheared value $Ra=1707.762$. A last modification of the 
critical Rayleigh number occurs because of lateral wall effects. As a 
result, the critical Rayleigh number increases with decreasing aspect 
ratio $\Gamma$ (lateral width over radial width). For example, for 
$\Gamma=5,2,1,0.5$, $Ra_c=1779,2013,2585,12113$. In most 
Taylor-Couette experiments, the aspect ratio is very large (typically 
above 8 or so). So the analog critical Rayleigh number is close to 
$1708$. However, many modern convection experiment (reaching very 
large Rayleigh numbers) deal with a rather small aspect ratio 
($\Gamma\sim 1$). This unfortunately limits the possibilities of 
direct comparisons between the Taylor-Couette experiments and the 
convective experiments to values close to the onset of instability. 
For larger values of Rayleigh numbers, we shall use extrapolations.\

\subsection{Angular momentum transfer}
A second interesting quantity in convection is the non-dimensional 
heat transfer $Nu=Hd/\kappa \Delta T$, where $H$ is the heat 
transfer. Via the analogy, the analog of this is the non-dimensional 
angular momentum transfer, which can be computed using the 
non-dimensional torque $G=T/\nu^2 L$, where $L$ is the cylinder 
length:
\EQA
Nu_*&\equiv&\frac{G}{G_{laminar}},\nonumber\\
&=&\frac{G}{R}\frac{(1+\eta)(1-\eta)^2}{4\pi\eta}.
\label{nuequiv}
\ENA
The normalization by $G_{laminar}$ ensures that in the laminar case, 
$Nu_*=1$, like in the convective analog.\

\subsubsection{Instability onset}
Theoretical \cite{SLB65} and experimental \cite{PlatLegr84} studies 
of convection near threshold lead to identification of two regimes 
just above the critical Rayleigh number:
\par for $\epsilon=\frac{Ra-Ra_c}{Ra_c}\le 1$, a linear regime in which
\EQ
\left(Nu-1\right)\frac{Ra}{Ra_c}= K_1 \epsilon.
\label{regimeonset1}
\EN
The constant $K_1$ depends on the Prandtl number. For $Pr=1$, it is 
$K_1\approx 1/0.7=1.43$ \cite{SLB65}.

\begin{figure}[hhh]
\includegraphics[clip=true,width=0.99\columnwidth]{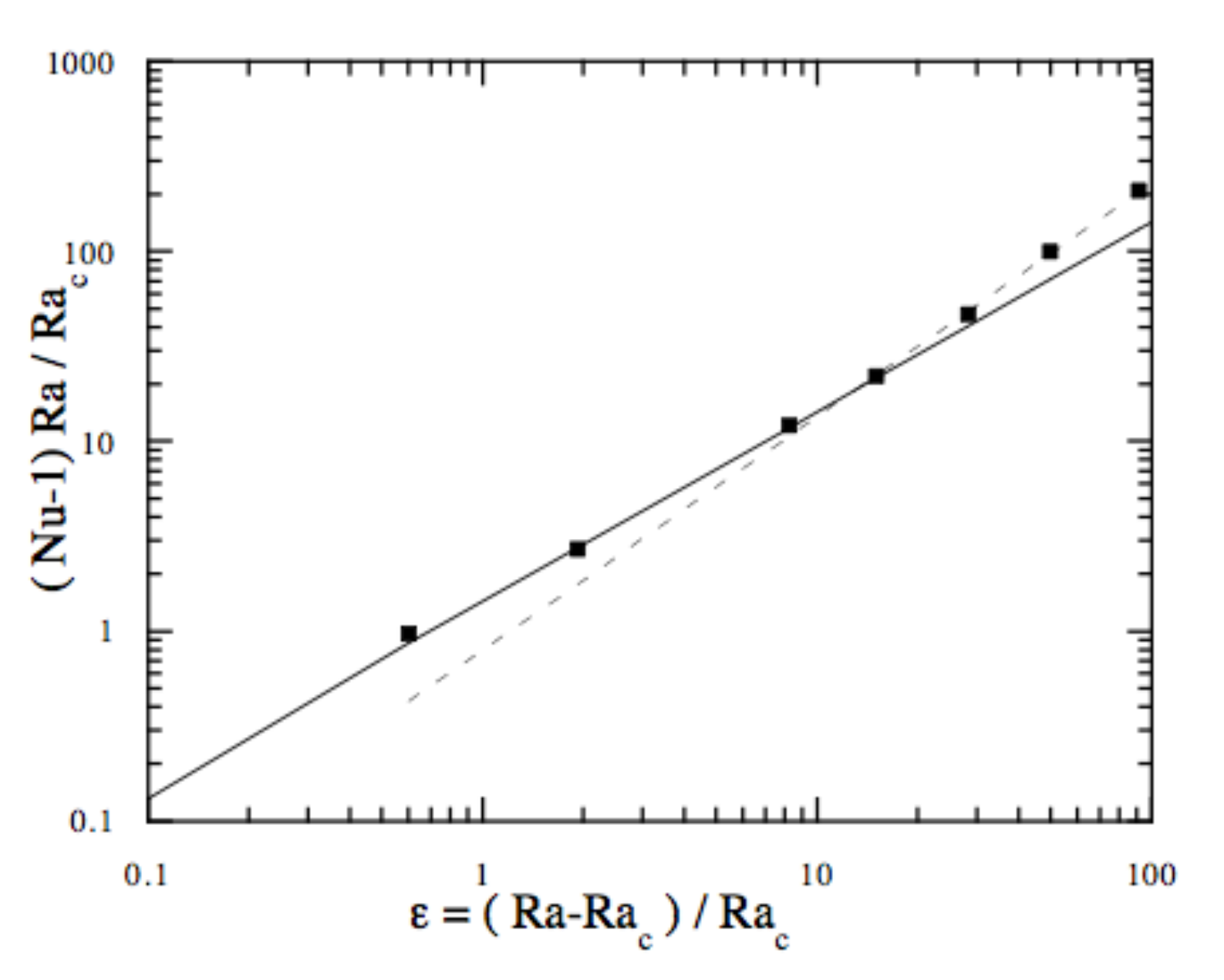}
\caption[]{ Comparison of the theoretical near instability onset 
behavior with the data of Wendt \protect\cite{Wendt33}.  The symbols 
are the experimental measurements. The two lines are the theoretical 
formula predicted by analogy with convection for $\epsilon<1$ 
($(Nu-1)Ra/Ra_c=1.43\epsilon$ and for $\epsilon>1$ 
($(Nu-1)Ra/Ra_c\sim\epsilon^{1.23}$.) In the latter case, the 
proportionality constant is not constrained by the analogy, and needs 
to be adjusted for a best fit.}
\label{fig:fig1}
\end{figure}

\par for larger $\epsilon$, a scaling regime in which \cite{PlatLegr84}
\EQ
\left(Nu-1\right)\frac{Ra}{Ra_c}= K_2 \epsilon^{1.23}.
\label{regimeonset2}
\EN
Here, $K_2$ is a constant which is not predicted by the theory.
In Fig. \ref{fig:fig1}, we show how the results of Wendt obtained with 
$\eta=0.935$ near the instability threshold compare with these two 
predictions. One sees that the linear regime is indeed obtained for 
$\epsilon\le 10$, while the scaling regime is obtained for larger 
values of $10<\epsilon<100$. Further from the threshold, one needs to 
compare with the turbulent theories of convection.

\subsection{The classical turbulent regimes}

Using (\ref{nuequiv}), we can single out some interesting regimes. 
Note that since $\kappa_*=\nu$, we are in the case of unit Prandtl 
number convection, i.e. for example convection in Helium ($Pr> 0.7$). 
In the classical theory of convection, one usually considers three 
regimes: a first one, labeled as "soft turbulence", in which $Nu\sim 
Ra^{1/3}$, $5\times 10^5<Ra<2\times 10^7$ \cite{HCL87}; then for 
$2\times 10^7<Ra< 10^{11}$, a "hard turbulence" regimes in which 
$Nu\sim Ra^{2/7}$ \cite{Cast89}; finally for $Ra>10^{11}$, a ultra-hard 
turbulence regime in which $Nu\sim Ra^{1/2}$ \cite{Krai62,CCCHCC97}. 
Using the analogy, we see that these three regimes translate into: 
for $707<R\eta(1-\eta)^{1/2}<4472$, $G\sim R^{5/3}$; for 
$4472<R\eta(1-\eta)^{1/2} <3\times 10^5$, $G\sim R^{11/7}$; for 
$(1-\eta)^{1/2}\eta R>3\times 10^5$, $G\sim R^2$. To evaluate the 
boundary between the two regimes, we have used (\ref{rayleighequiv}) 
at $\eta=1$.\

The first regime has been predicted by \cite{KLSM84,BarcBrin84} using 
marginal stability analysis. The third regime can be derived from 
Kolmogorov type arguments (see e.g. \cite{LFS92}). It also 
corresponds to some upper-bound in the angular momentum transport 
\cite{DoerConst92}. The intermediate regime is new, and leads to a 
scaling exponent of $1.57$. Experimentally, some of this scaling 
regimes have been reported, but not in the same sequence: in his 
experiments with $0.680<\eta<0.935$, Wendt \cite{Wendt33} reports a 
scaling exponent of $1.5$ for $400<R<10^4$, followed by a scaling 
exponent $1.7$ for $10^4<R<10^5$. In more recent experiments, Lathrop 
et al \cite{LFS92} measure  a ``local"  exponents $d\ln(G)/d\ln(R)$ 
which varies continuously from $1.2$ to $1.9$, with a transition at 
$R\sim 1.3\times 10^4$ (for $\eta=0.7246$). This transition was later 
found to correspond to a modification of coherent structures in the 
flow \cite{LewiSwin99}. Remarkably enough, the analog Rayleigh number 
characterizing this transition is $Ra_*=2\times 10^7$, like in 
convection. We may therefore interpret it as the boundary between 
"soft" and "hard" turbulence.  However, the scaling reported in 
\cite{LFS92} does not seem to fully correspond to the soft and hard 
turbulence scaling.
In the sequel, we wish therefore to explore a new possibility, based 
on logarithmic corrections to scaling.

\subsection{The logarithmic turbulent regimes}
\begin{figure}[hhh]
\includegraphics[clip=true,width=0.99\columnwidth]{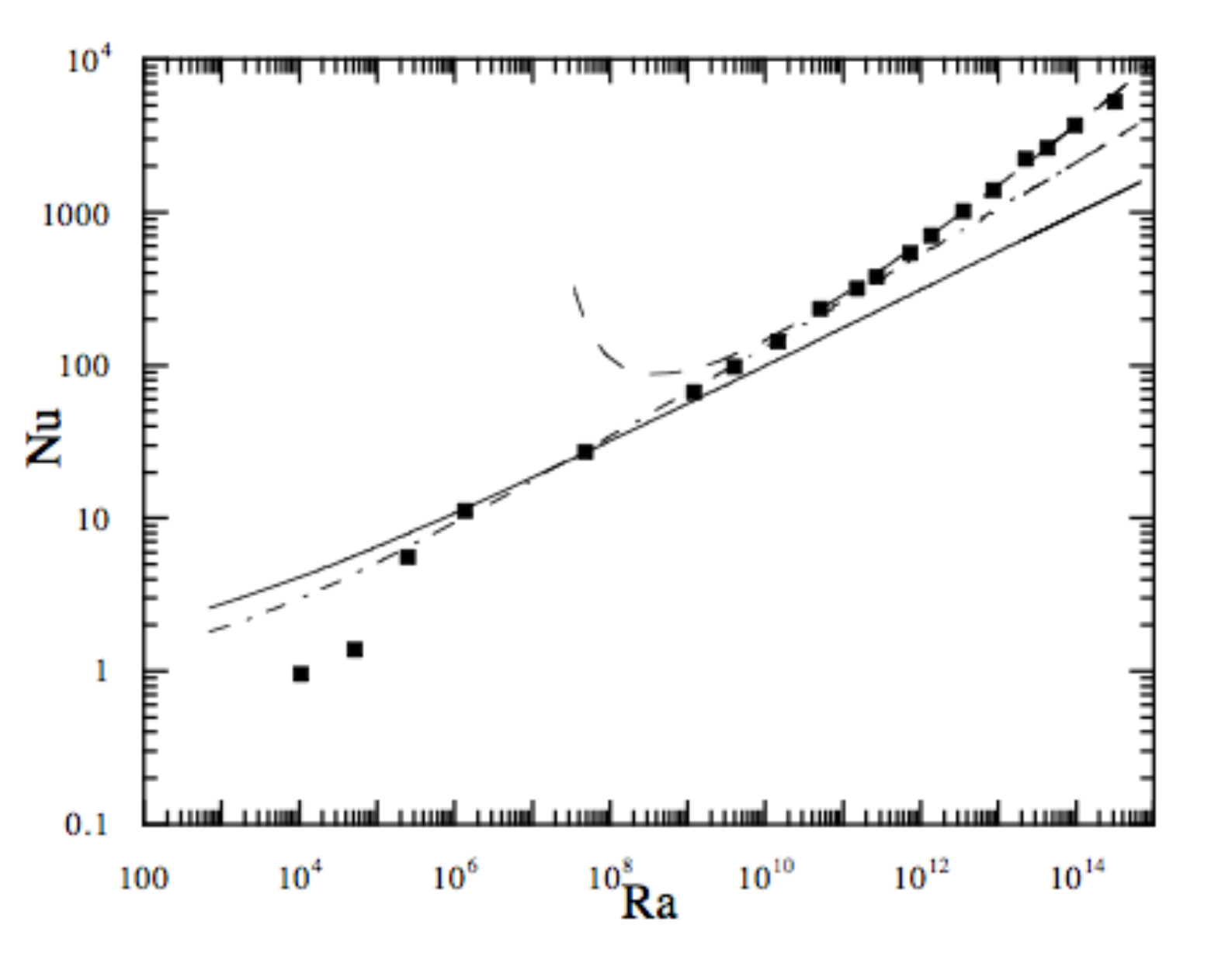}
\caption[]{ Illustration of the three scaling regimes found in 
convection in Helium for Nusselt vs Rayleigh. The symbols are 
experimental measurements by
\protect\cite{CCCHCC97}. The lines are theoretical prediction by 
\protect\cite{Dubr01} using an analytical  model of turbulent 
convection. "Soft" turbulence regime(mean flow dominated): power law 
$Nu\sim Ra^{1/4}$ (full line); "Hard" turbulence regime: (velocity 
fluctuation dominated) $Nu\sim Ra^{1/3}/(\ln(Ra))^{2/3}$ (dotted 
line); "Ultra-hard" turbulent regime: (temperature fluctuations 
dominated) $Nu\sim Ra^{1/2}/(\ln(Ra))^{3/2}$ (dashed line) }
\label{fig:fig2}
\end{figure}

The observation by \cite{LFS92} that no scaling prevails  for the 
angular momentum transport has in fact its counter-part for the heat 
transport in convection \cite{GrosLohs00}. In a recent work, we used 
the turbulent model to analytically compute the heat transport in a 
convective cell. At $Pr=1$, we found 3 different regimes: at low 
Rayleigh number, the dissipation is dominated by the mean flow, and 
$Nu=K_1 Ra^{1/4}Pr^{-1/12}$; at larger Rayleigh number, the kinetic 
energy dissipation starts being dominated by velocity fluctuations, 
and the heat transport becomes $Nu Pr^{1/9}=K_2 Ra^{1/3}/\ln(Ra 
Pr^{2/3}/20)^{2/3}$. Finally at very large Rayleigh number, the heat 
dissipation becomes also dominated by (heat) fluctuations, and $Nu= 
K_3 Ra^{1/2}/\ln( Ra/Ra_c)^{3/2}$. Fig. \ref{fig:fig2} shows the illustration of 
these 3 regimes in a Helium experiment of \cite{CCCHCC97}, with the 
three fits corresponding to these 3 regimes. From this graph, we 
obtain $K_1=0.31 $, $K_2=0.45$, $K_3=0.023$ and $Ra_c=2\times 10^7$, 
for an aspect ratio of $0.5$. These constants tend to decrease 
slightly for larger aspect ratio by an asymptotic factor of about 
$0.75$ (at $Ra=10^8$, see table 1 of \cite{CCCHCC97}). The small 
aspect ratio of the experiment also increases the critical Rayleigh 
for instability from near $1708$ to near $4\times 10^4$. The boundary 
between the regime 1 and 2 lies at $Ra=1.5\times 10^8$. It is 
characterized by a change in the temperature statistics, going from 
nearly Gaussian to exponential. The boundary between the regime 2 and 
3 is somehow ill defined, and lies between $Ra=2\times10^{10}$ and 
$Ra=10^{11}$. Note than in a similar experiment, ran by another group, the third 
regime was not detected, even at $Ra=10^{15}$ \cite{NSSD00}. 
The reason of this difference is not yet known. A possibility would be that different boundary conditions may or may not allow the growth of the temperature perturbation, thereby favoring or inhibiting this last regime \cite{Dubr01}.\

The translation of the three logarithmic regimes using the analogy 
gives a priori three possible regimes in the Taylor-Couette 
experiments.

In the regime 1, we get:
\EQ
G=K_4\frac{\eta^{3/2}}{(1-\eta)^{7/4}}R^{3/2}.
\label{regime1taylor}
\EN
In the regime 2, we get:
\EQ
G=K_5\frac{\eta^{2/3}}{(1-\eta)^{5/3}}\frac{R^{5/3}}{\ln[\eta^2(1-\eta 
)R^2/K_6]^{2/3}},
\label{regime2taylor}
\EN
while in the regime 3, we get:
\EQ
G=K_7\frac{\eta^{2}}{(1-\eta)^{3/2}}\frac{R^{2}}{\ln[\eta^2(1-\eta)R^2 
/K_8]^{3/2}},
\label{regime3taylor}
\EN
In these expressions, we have introduced 5 unknown coefficients, 
which a priori depend on the aspect ratio. Since there is no 
available large Rayleigh number large aspect ratio convection 
experiments, we shall extrapolate or fit these coefficients by 
comparison with Taylor-Couette data.\

\subsection{Comparison with experiments}
For this, we use torque measurements from Wendt \cite{Wendt33} and 
\cite{LFS92,LewiSwin99}.
The regime 1 should be observed at rather moderate Reynolds numbers. 
Therefore, it explains very well the old measurements by Wendt 
\cite{Wendt33} who found the same exact dependence in $\eta$ and $R$ 
for $400<R<10^4$, and with  a prefactor of $K_4=1.45$. The analogy 
with convection predicts that $K_4=2\pi K_1$. The small aspect ratio 
convective experiment extrapolated at large aspect ratio gives 
$K_1=0.75\times 0.31$, which translates into $K_4=1.46$. This is in 
very good agreement with the prefactor measured by Wendt.\

\begin{figure}[hhh]
\includegraphics[clip=true,width=0.99\columnwidth]{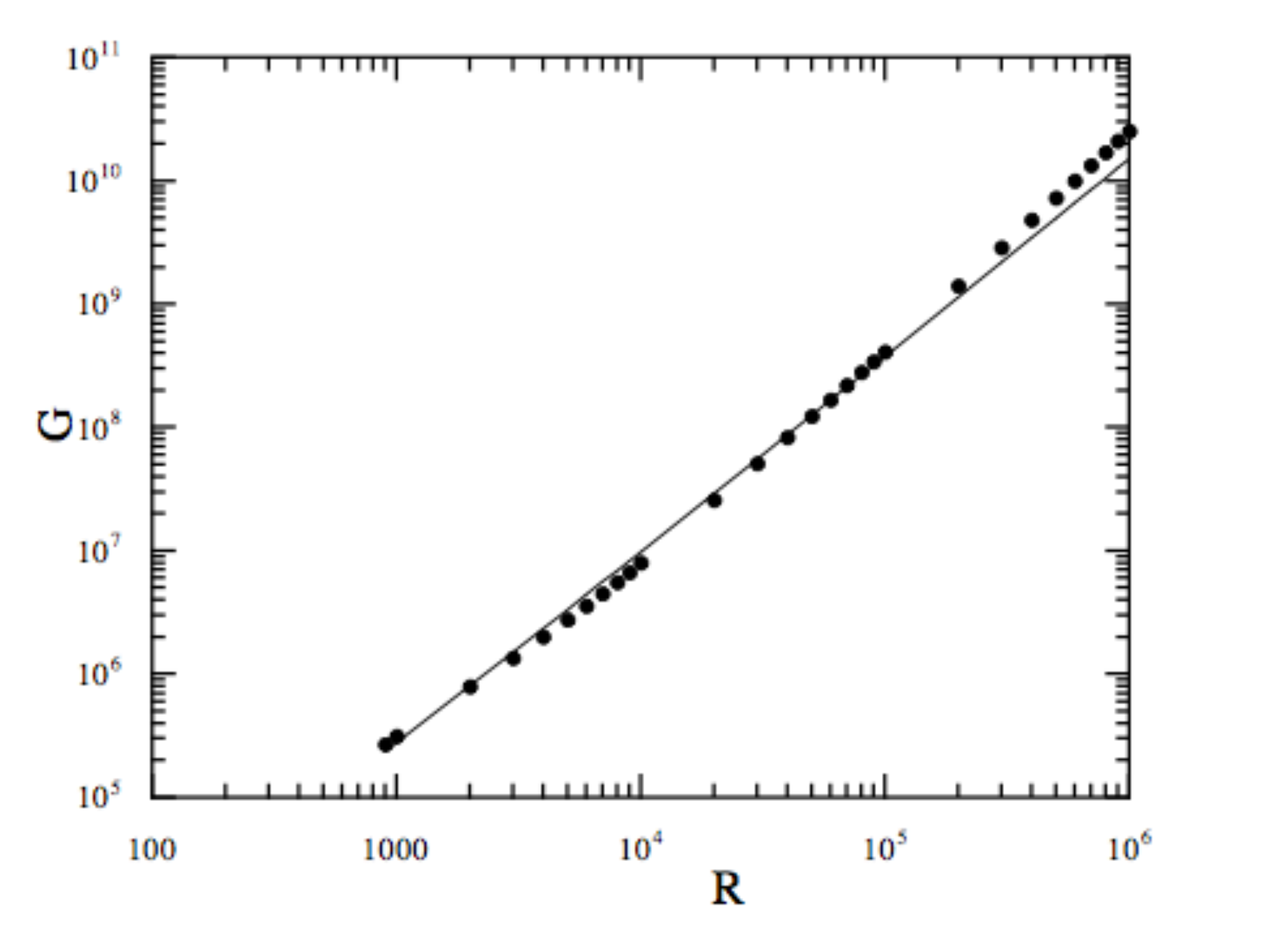}
\caption[]{Torque vs Reynolds in Taylor-Couette experiments. The 
symbols are  the data of 
\protect\cite{LewiSwin99}. The line is the theoretical formula 
obtained in the hard turbulence regime and computed using the analogy 
with convection $G=A R^{5/3}/(\ln( R/B))^{2/3}$. The two constants 
$A$ and $B$ are not fitted to the data, but are analytically computed 
using the analogy with convection.}
\label{fig:fig3}
\end{figure}

\begin{figure}[hhh]\includegraphics[clip=true,width=0.99\columnwidth]{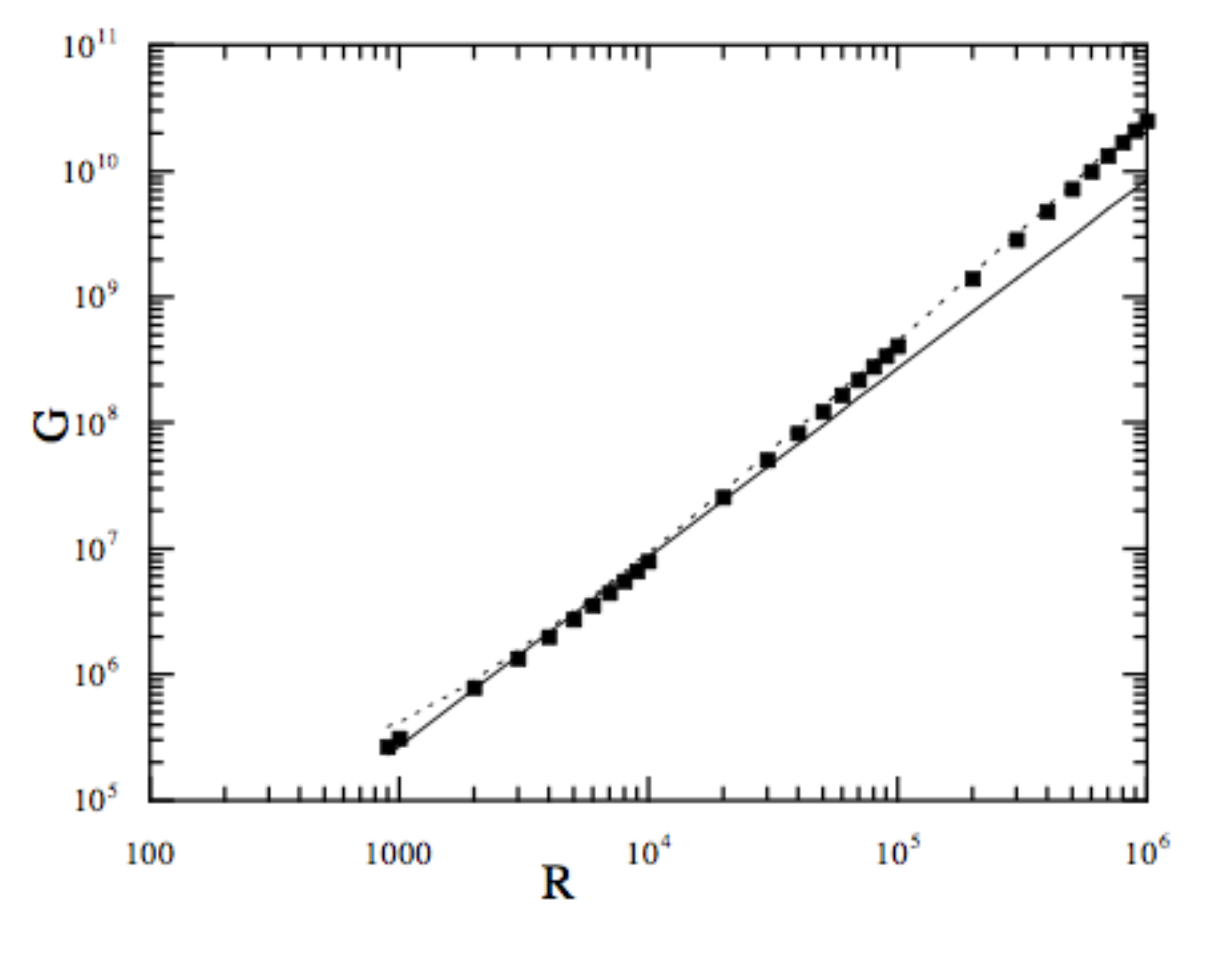}
\caption[]{Torque vs Reynolds in Taylor-Couette experiments. The 
symbols are  the data of 
\protect\cite{LewiSwin99}. The lines are the theoretical formula 
obtained in the soft and ultra-hard turbulence regimes and computed 
using the analogy with convection. Soft turbulence eq. 
(\ref{regime1taylor}) (full line); ultra-hard turbulence eq. 
(\ref{regime3taylor}) (dotted line). In the former case, all the 
constants are analytically computed using the analogy. In the latter 
case, we have seek the best adjustment with data by adjusting the two 
constants.
}
\label{fig:fig4}
\end{figure}

\begin{figure}[hhh]
\includegraphics[clip=true,width=0.99\columnwidth]{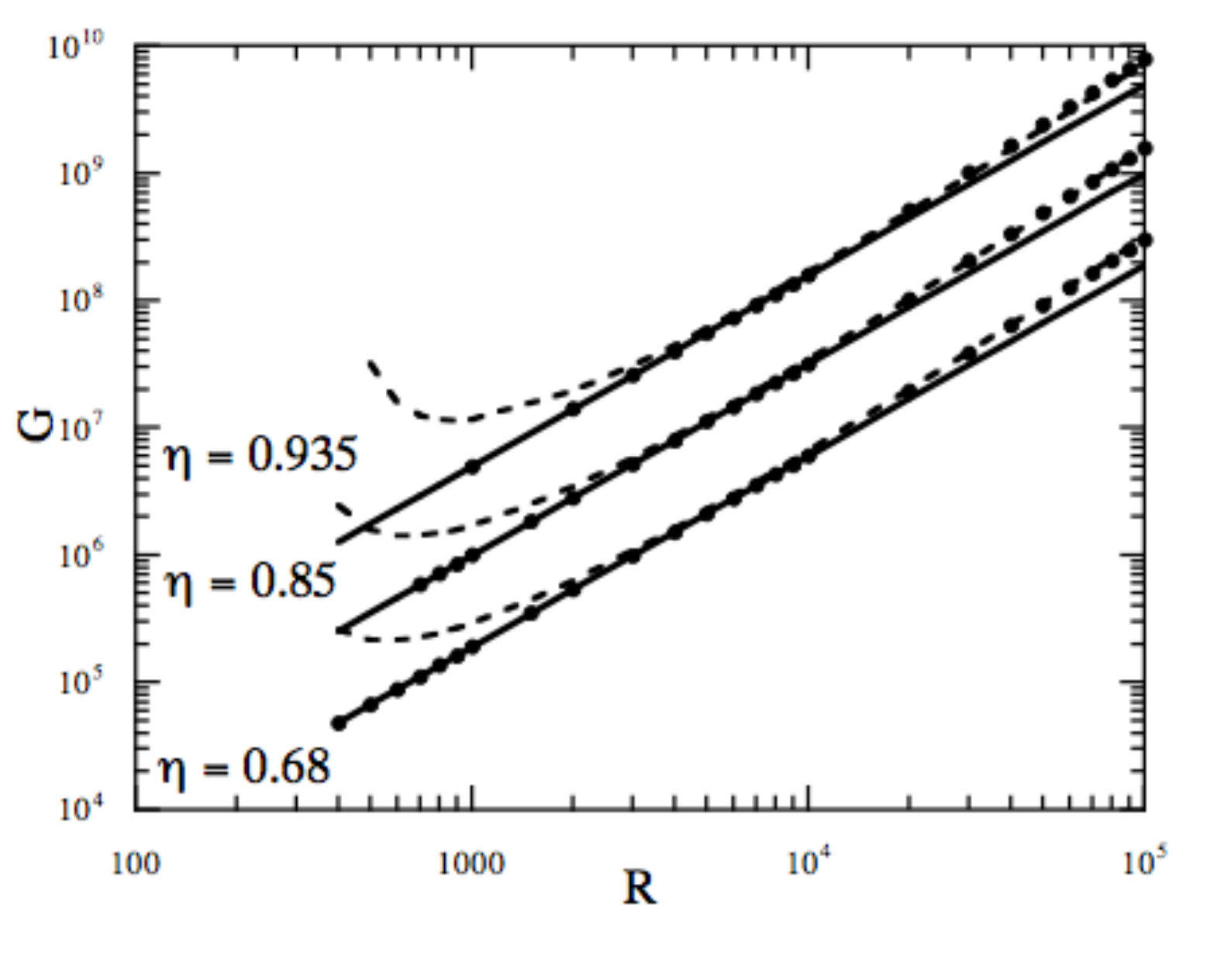}
\caption[]{Torque vs Reynolds in Taylor-Couette experiments for 
different gap widths $\eta=0.68$, $\eta=0.85$ and $\eta=0.935$. The 
symbols are the data of 
\protect\cite{Wendt33}. The lines are the theoretical formula 
obtained in the soft and ultra-hard turbulence regimes and computed 
using the analogy with convection. Soft turbulence eq. 
(\ref{regime1taylor}) (full line); ultra-hard turbulence eq. 
(\ref{regime3taylor}) (dotted line). There is no adjustable parameter 
in this comparison, all the constants being fixed either by the 
analogy with convection, or by the comparison with the data of 
\protect\cite{LewiSwin99}.}
\label{fig:fig5}
\end{figure}

The second regime predicts torque varying more slowly than $R^{5/3}$. 
It could therefore only marginally explain the second regime observed 
by Wendt, for $R>10^4$, in which $G\sim R^{1.7}$. However, it could 
explain the regime obtained by \cite{LFS92,LewiSwin99} for $R\sim 
10^4$, in which a continuously varying scaling exponent was obtained. 
This is shown in Fig. \ref{fig:fig3}, where the fit to the data of 
\cite{LewiSwin99} is compared with the theoretical formula 
(\ref{regime2taylor}). The comparison is made using coefficients 
extrapolated from the small aspect ratio convection experiment: 
$K_6=20$, $K_5= 2\pi K_2$ with $K_2=0.75\times 0.45$. It may happen 
however that this regime 2 does not exist in Taylor-Couette 
experiments. Indeed, since the temperature analog is related to the 
velocity, it might be impossible to excite velocity fluctuations 
without exciting pseudo-temperature fluctuations. This would mean a 
direct transition from regime 1 (mean flow dominated) to regime 3 
(fluctuation dominated). This possibility is explored in Fig. \ref{fig:fig4}, 
where we show the best fit of the measurements of Lewis and Swinney, 
with formula (\ref{regime3taylor}). This fit uses $K_7=0.50$ and 
$K_8=10^4$. Notice the big difference between these constants and 
their extrapolation from the convective case $K_7=0.145$ and 
$K_8=2\times 10^7$. This may reflect the sensitivity to boundary 
conditions of the regime 3. Note however that the fit is excellent 
from $R=10^3$ up to $R=10^6$. Below $R=10^4$, the regime 1 with 
$K_2=1.46$ fits the data very well also. As a last check, we have 
compared this regime 3 with the constant fitted for Lewis and 
Swinney's data, to the data of Wendt. The result is 
shown in Fig. \ref{fig:fig5}, for 3 different gap $\eta=0.68,0.85,0.935$. The 
agreement is excellent.

\subsection{Velocity fluctuations}
The analogy can also be used to predict the behavior of velocity 
fluctuations. In \cite{LewiSwin99}, the azimuthal turbulent intensity 
$i_\theta=\sqrt{<u_\theta^2>}/U_\theta$ was measured at midgap with 
hot film probes. Above $1\times 10^4$, a fit yields
\EQ
i_\theta=0.10 R^{-0.125}.
\label{velocityfluct}
\EN
Using the analogy, this intensity is related to the temperature 
fluctuations at mid-gap, in the ultra hard turbulent regime (regime 
3). The total analog temperature fluctuation in fact also includes 
vertical velocity fluctuations (see Table 1). In an axisymmetric 
turbulence, one could therefore expect that the turbulent intensity 
measured by Lewis and Swinney is proportional to the temperature 
analog. Recent measurements of this quantity at Rayleigh number up to 
$Ra=10^{15}$ have been measured by \cite{NSSD00} in a low aspect 
ratio Helium experiment. They found $\theta/\Delta T=0.37 
Ra^{-0.145}$, but this was obtained in a regime where the Nusselt 
number varies like in regime 2 (velocity fluctuation dominate but NOT 
temperature fluctuation). Using the analogy, this would translate 
into a regime where $i_\theta\sim R^{-0.29}$, in clear contradiction 
with the data of Lewis and Swinney, see Fig. 6. This might therefore 
be another proof of the absence of the regime 2 in Taylor-Couette 
experiment.\

\begin{figure}[hhh]
\includegraphics[clip=true,width=0.99\columnwidth]{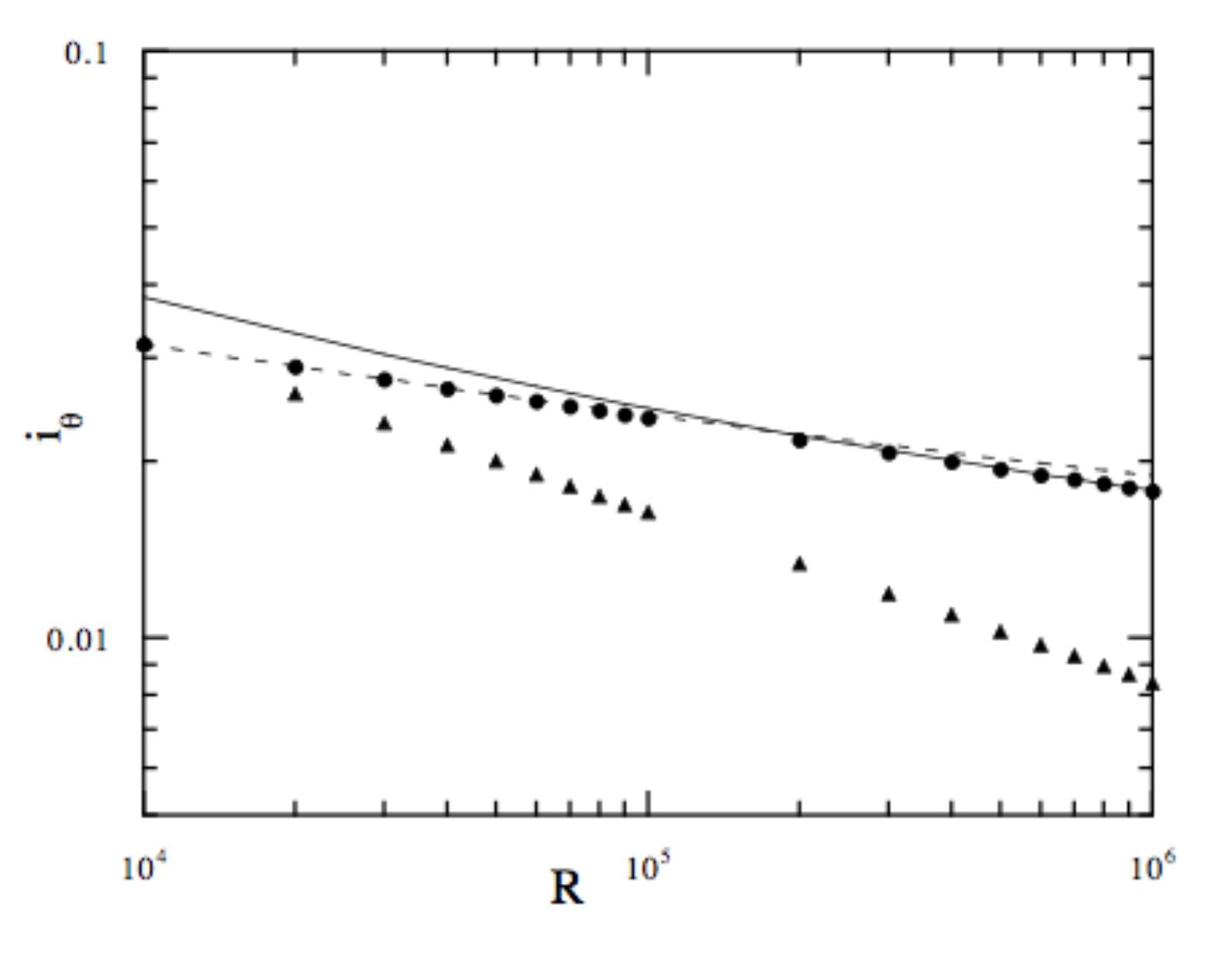}
\caption{ Azimuthal velocity fluctuations in Taylor-Couette flow. The 
circles are the power-law fits of experimental measurements by
\protect\cite{LewiSwin99}. The triangles are the power-law fit to the 
temperature fluctuations (analog of azimuthal velocities) in Helium 
by \protect\cite{NSSD00}. The line is the prediction obtained with 
the analog of the free-convective regime \protect\cite{DearWill67}. 
The dotted line is the theoretical formula predicted by our model eq. 
(\ref{predimodel}).}
\label{fig:fig6}
\end{figure}

Unfortunately, we are not aware of temperature measurements in 
convective turbulence in the ultra-hard regime. In previous analysis 
of temperature fluctuations in the atmospheric boundary layer, 
Deardoff and Willis \cite{DearWill67} showed that temperature 
fluctuations follow  the free convection regime
\EQ
\frac{\theta}{\Delta T}\propto \frac{Nu}{(Pr Ra Nu)^{1/3}},
\label{freeconv}
\EN
where the proportionality constant is of the order 1. Fig. 6 shows 
the application of this scaling to the data of Lewis and Swinney, 
where the analogy was used to translate torque and Reynolds into 
Nusselt and Rayleigh. The best agreement with the experimental fit of 
Lewis and Swinney is obtained for a prefactor $1.8$.
We can also compare the results with the theoretical prediction given 
by the convective model. In this model \cite{Dubr01}, the temperature 
fluctuations in the boundary layer obey:
\EQ
\frac{<\theta'^2>}{\Delta T^2}= 
\lambda_u\frac{Nu^{5/2}}{(RaPr)^{1/2}}\frac{\sqrt{1+(z/\lambda_u)^2}}{ 
1+(zNu)^2},
\label{fluctutheta}
\EN
where $\lambda_u$ is the height of the viscous velocity layer.
The value at the height of the boundary layer is obtained for 
$z=\lambda_{BL}$, the size of the boundary layer, which was shown to 
vary like $\lambda_{BL}\sim (Ra Nu)^{-1/8}/\sqrt{\ln(Ra Nu)}$.
Assuming that the value at mid-gap equals this maximal value, we obtain:
\EQ
\frac{\sqrt{<\theta'^2>}}{\Delta 
T}=K_{10}\frac{Nu^{5/16}}{Ra^{3/16}}\left(\ln(Ra 
Nu/K_{11})\right)^{1/4}.
\label{predimodel}
\EN
This prediction is shown in Fig. \ref{fig:fig6}, using a fitted prefactors of 
$K_{10}=0.16$ and $K_{11}=1$. It is in very good agreement with the 
experimental fit of Lewis and Swinney.

\section{Conclusion}
In this paper, we have shown how a well-known analogy between 
stratified and rotating shear flows, for axi-symmetric perturbations, can be extended into the 
turbulent regime. Assuming predominance of the axi-symmetric perturbation in the turbulence dynamics , we used this analogy in the unstable case (analogy between convection and centrifugally unstable Taylor-Couette flow) to predict the scaling of the momentum 
transfer and velocity fluctuations. Our prediction is that at low 
Reynolds number, the non-dimensional torque follows
(\ref{regime1taylor}) while at $R>10^4$, it follows 
(\ref{regime3taylor}). The analogy can also be used to discriminate 
between theories about Taylor-Couette turbulent quantities. For 
example, we have shown that the "classical" $Nu\sim Ra^{1/3}$ regime, 
translate into a $G\sim Ra^{5/3}$ in the Taylor-Couette flow (both 
being unobserved experimentally at large Rayleigh or Reynolds 
number).\

The analogy also sheds new light on the recent theory of Eckhardt et al 
\cite{EGL00}. It predicts a dependence: $G=c_1 Re^{3/2+5\xi/2}+c_2 Re^{2+3\xi}$, where $\xi=-0.051$ is a parameter which has been adjusted to a best fit. When translated using this analogy, this formula would give in the convective case: $Nu=c_1 Ra^{1/4+5\xi/4}+c_2 
Ra^{1/2+3\xi/2}$. This has to be compared with the theoretical 
prediction of Grossman and Lohse \cite{GrosLohs00}, made using the 
same theory, which leads to $Nu=c_1 Ra^{1/4}+c_2 Ra^{1/3}$. Clearly, there is no value of $\xi$ which can reconciliate the two formulae. It would therefore be interesting to see whether the analog of the Grossman and Lohse formula, namely:
$G=c_1=Re^{3/2}+c_2 Re^{5/3}$ would not fit the data equally well than the Eckhardt et al formula. This would reduce the number of unknown parameter by one.\

It would now be interesting to study in more details consequences of the analogy in the stable case (i.e. stably stratified flow v.s. centrifugally stable flow). 
There are many observational, numerical and experimental results in the case of stably stratified flows. However, their counter part in the centrifugally stable rotating case is presently missing. Recent experiments by Richard et al \cite{Richardthese} performed on flows between counter-rotating cylinders could help filling this gap.\

Finally, the analogy is of great interest for astrophysical and 
geophysical applications. In astrophysics, for example, many objects 
are differentially rotating, and are characterized by very large 
Reynolds number. These Reynolds numbers cannot be reached in 
laboratory experiments. On the other hand, we have at our disposal a 
natural high Rayleigh (and Reynolds number) laboratory of stratified 
turbulence: the atmospheric boundary layer. We believe that we could 
use all the data collected in our atmosphere to get great insight 
about large Reynolds number behavior of rotating, astrophysical shear 
flows, using the analogy sketched in the present paper.\

We thank Fran\c cois Daviaud for comments on the manuscript.

\begin{table}
\begin{tabular}{c|c}
stratified shear flow &rotating shear flow\\
\hline
$z$ &$r$\\
$x$ &$\theta$\\
$\partial_z U$ &$r\partial_r \Omega$\\
$\beta g$ &$ 2\frac{\Omega}{r}\sin^2\phi$\\
$\partial_z\Theta$ &$\frac{1}{r}\partial_r (r^2\Omega)$\\
$w$ &$u$\\
$\theta$ &$(rv-w\cot\phi)$\\
\end{tabular}
\label{table1}
\caption{The detailed analogy between startified and rotating shear flow. The notations for the stratified case are from \protect\cite{DLS02}.}
\end{table}

\end{document}